\documentclass{cargese}
\usepackage{graphicx}
\usepackage{amssymb}

\let\footnote\savefootnote
\let\footnotetext\savefootnotetext 
\let\footnoterule\savefootnoterule 
\normallatexbib

\begin{document}
\articletitle{Integrability of Superconformal Field Theory 
and SUSY N=1 KdV}
\author{Anton M. Zeitlin}
\affil{Department of High Energy Physics, Physics Faculty,\\ 
St. Petersburg State University, Ul'yanovskaja 1, Petrodvoretz, \\
St. Petersburg, 198904, Russia}
\email{zam@math.ipme.ru, http://www.ipme.ru/zam.html}

\begin{abstract}
The quantum SUSY N=1 hierarchy based on $sl(2|1)^{(2)}$ twisted affine 
superalgebra is considered. The construction of the corresponding 
Baxter's Q-operators 
and fusion relations is outlined. The relation with the superconformal 
field theory is discussed.   
\end{abstract}

One of the most famous integrable systems (IS) is the Korteweg-de Vries 
hierarchy. It is related with the superconformal field theory because its Poisson 
brackets give the Virasoro algebra and the involutive family of integrals 
of motion (IM) providing the integrability of the conformal field theory 
(CFT). Since the late 1980s the supersymmetric and fermionic extensions of 
the KdV system have been known (see e.g. \cite{sKdV}, \cite{plb1}, \cite{plb2} and references therein), 
which in turn are related with superconformal field theory (SCFT). During the 
following years they were extensively studied on both the classical 
and the quantum level.\\
\hspace*{5mm}However, up to the present nobody has applied the most successful 
method in the theory of integrable systems, the so-called quantum inverse 
scattering method (QISM) to these IS. In this short paper we demonstrate 
some algebraic tools giving possibility to study SUSY N=1 KdV via QISM.  
\section{RTT-relation}   
The SUSY N=1 KdV model is related to the following L-operator:
$$
\mathcal{L}_F=D_{u,\theta} 
-D_{u,\theta}\Phi h_{\alpha}-(e_{{\delta-\alpha}}+ e_{{\alpha}}),
$$
where $h_{\alpha}$, $e_{{\delta-\alpha}}\equiv e_{\alpha_0}$, 
$e_{{\alpha}}$ are the 
Chevalley generators of twisted affine Lie superalgebra 
$sl(2|1)^{(2)}\cong osp(2|2)^{(2)}\cong C(2)^{(2)}$, 
$D_{u,\theta} =\partial_\theta + \theta \partial_u$ 
is a superderivative, the variable 
$u$ lies on a cylinder of circumference $2\pi$, $\theta$ 
is  a  Grassmann  variable, $\Phi(u,\theta)=\phi(u) - 
\frac{i} {\sqrt{2}}\theta\xi(u)$ 
is a bosonic superfield with the following Poisson brackets: $
\{D_{u,\theta}\Phi(u,\theta), D_{u',\theta'}\Phi(u',\theta')\}=  
D_{u,\theta}(\delta(u-u')(\theta-\theta'))$.
Making a gauge transformation of the L-operator we obtain a new superfield 
$\mathcal{U}(u,\theta)\equiv
D_{u,\theta}\Phi(u,\theta)\partial_u\Phi(u,\theta)-D_{u,\theta}^3
\Phi(u,\theta)$ =$-\theta U(u)-i\alpha(u)/\sqrt{2}$, 
where $U$ and $\alpha$ generate the superconformal algebra under the Poisson 
brackets:
\begin{eqnarray}
\{U(u),U(v)\}&=&
 \delta'''(u-v)+2U'(u)\delta(u-v)+4U(u)\delta'(u-v),\nonumber\\
\{U(u),\alpha(v)\}&=&
 3\alpha(u)\delta'(u-v) + \alpha'(u)\delta(u-v),\nonumber\\
\{\alpha(u),\alpha(v)\}&=&
 2\delta''(u-v)+2U(u)\delta(u-v)\nonumber.
\end{eqnarray} 
 The SUSY N=1 KdV system has an infinite number of conservation laws
 and the first nontrivial one gives the SUSY N=1 KdV equation: 
$
\mathcal{U}_t=-\mathcal{U}_{uuu}+3(\mathcal{U} D_{u,\theta}\mathcal{U})_u
.
$
The integrals of motion are generated by the logarithm of the supertrace 
of the corresponding monodromy matrix, which has the following form:
$$
\mathbf{M}^{(cl)}=e^{2\pi i ph_{\alpha_1}}
P\exp\int_0^{2\pi} d u\Big(\frac{i}{\sqrt{2}}
\xi(u)e^{-\phi(u)}e_{\alpha_1}
$$
$$
-\frac{i}{\sqrt{2}}
\xi(u)e^{\phi(u)}e_{\alpha_0}
-e^2_{\alpha_1}e^{-2\phi(u)}-
e^2_{\alpha_0}e^{2\phi(u)}-[e_{\alpha_1},e_{\alpha_0}]
\Big).
$$
Its quantum generalization can be represented in the quantum P-expo\-nen\-tial 
form 
(for the explanation of this notion see below and \cite{plb2} for details):
$$
\mathbf{M}^{(q)}=e^{2\pi iPh_{\alpha_1}}Pexp^{(q)}\int^{2\pi}_{0}d u
(W_{-}(u)e_{\alpha_1} +W_{+}(u)e_{\alpha_0}).\nonumber
$$
Vertex operators $W_{\pm}$ are defined in the following way 
$W_{\pm}(u)=\int d \theta :e^{\pm\Phi(u,\theta)}:= \mp \frac{i}{\sqrt{2}}
\xi(u):e^{\pm\phi(u)}:$.
The universal R-matrix with the lower Borel subalgebra represented 
by $(q^{-1}-q)^{-1}\int_{0}^{2\pi}d u W_{\pm}(u)$ is equal to
$\mathbf{L}=e^{-\pi i P h_{\alpha_{1}}}\mathbf{M}^{(q)}$.
Due to this fact $\mathbf{L}$ satisfies the RTT-relation: 
$$
\mathbf{R}_{ss'}
\Big(\mathbf{L}_s\otimes \mathbf{I}\Big)\Big(\mathbf{I}
\otimes \mathbf{L}_{s'}\Big)
=(\mathbf{I}\otimes \mathbf{L}_{s'}\Big)
\Big(\mathbf{L}_s\otimes \mathbf{I}\Big)\mathbf{R}_{ss'},\nonumber
$$
where $s$, $s'$  mean that the corresponding object is considered 
in some representation 
of $C_q(2)^{(2)}$. 
Thus the supertraces of the monodromy matrix (``transfer matrices'') 
$\mathbf{t}_s=str\mathbf{M}_s$ commute, providing the quantum integrability.
It is very useful to consider the evaluation representations of 
$C_q(2)^{(2)}$, $\rho_s(\lambda)$, where now the symbol $s$ means integer and 
half-integer 
numbers. Denoting $\rho_s(\lambda)(\mathbf{M})$ as $\mathbf{M}_s(\lambda)$ 
we find that $\mathbf{t}_s(\lambda)=str\mathbf{M}_s(\lambda)$ commute:
$ 
[\mathbf{t}_s(\lambda),\mathbf{t}_{s'}(\mu)]=0.\nonumber
$
The expansion of $\log(\mathbf{t}_{\frac{1}{2}}(\lambda))$ in $\lambda$ 
(the transfer matrix in the fundamental 3-dimensional representation) is believed to 
give us as coefficients the local IM, the quantum counterparts of the 
mentioned IM of SUSY $N=1$ KdV.
\section{The Q-operator}
\hspace*{5mm} Using the super q-oscillator representations of the upper Borel
 subalgebra of the quantum affine superalgebra $C_{q}(2)^{(2)}$ we define 
the $\mathbf{Q}_{\pm}$ operators (see \cite{npb} for details). 
The transfer-matrices in different evaluation representations can be expressed in such a way:
$$
2cos(\pi P)\mathbf{t}_s(\lambda)=\mathbf{Q}_{+}(q^{s+\frac{1}{4}}\lambda)\mathbf{Q}_{-}(q^{-s-\frac{1}{4}}\lambda)+\mathbf{Q}_{+}(q^{-s-\frac{1}{4}}\lambda)\mathbf{Q}_{-}(q^{s+\frac{1}{4}}\lambda),\nonumber
$$
where $s$ runs over integer and half-integer nonnegative numbers. $\mathbf{Q}_{\pm}$ operators satisfy quantum super-Wronskian relation:
$$
2cos(\pi P)=\mathbf{Q}_{+}(q^{\frac{1}{4}}\lambda)\mathbf{Q}_{-}(q^{-\frac{1}{4}}\lambda)+\mathbf{Q}_{+}(q^{-\frac{1}{4}}\lambda)\mathbf{Q}_{-}(q^{\frac{1}{4}}\lambda).\nonumber
$$ 
One should note, that we use only $4s+1$-dimensional 
``$osp(1|2)$-induced'' representations (sometimes called atypical) of $C(2)^{(2)}$. It allows, however, to construct the fusion relations, see below.
To construct the relations like Baxter's ones we introduce additional 
``quarter''-operators, constructed ``by hands'' from the 
$\mathbf{Q}$-operators:
$$
2cos(\pi P)\mathbf{t}_{\frac{k}{4}}(\lambda)=\mathbf{Q}_{+}(q^{\frac{k}{4}+\frac{1}{4}}\lambda)\mathbf{Q}_{-}(q^{-\frac{k}{4}-\frac{1}{4}}\lambda)-
\mathbf{Q}_{+}(q^{-\frac{k}{4}-\frac{1}{4}}\lambda)
\mathbf{Q}_{-}(q^{\frac{k}{4}+\frac{1}{4}}\lambda)
$$
for odd integer $k$.
 The Baxter's relations are:
$$
\mathbf{t}_{\frac{1}{4}}(\lambda)\mathbf{Q}_{\pm}(\lambda)=\pm \mathbf{Q}_{\pm}(q^{\frac{1}{2}}\lambda)\mp\mathbf{Q}_{\pm}(q^{-\frac{1}{2}}\lambda),
$$
$$
\mathbf{t}_{\frac{1}{2}}(q^{\frac{1}{4}}\lambda)\mathbf{Q}_{\pm}(\lambda)=\mathbf{t}_{\frac{1}{4}}(q^{\frac{1}{2}}\lambda)\mathbf{Q}_{\pm}(q^{-\frac{1}{2}}\lambda)+\mathbf{Q}_{\pm}(q\lambda).
$$
The fusion relations have the following form very similar to the $A_1^{(1)}$ case:
$$
\mathbf{t}_{j}(q^{\frac{1}{4}}\lambda)\mathbf{t}_{j}(q^{-\frac{1}{4}}\lambda)=
\mathbf{t}_{j+\frac{1}{4}}(\lambda)\mathbf{t}_{j-\frac{1}{4}}(\lambda)+(-1)^{4j}.  
$$
But they are only ``fusion-like'' because the ``quarter''-operators do not 
seem to correspond to any representation of $C(2)^{(2)}$. 
The truncation of these relations for different values of $q$, 
being the root of unity: $q^N=\pm1$, $N\in\mathbb{Z}$, $N>0$ has the following 
form:
$$
\mathbf{t}_{\frac{N}{2}}(\lambda)+\mathbf{t}_{\frac{N}{2}-
\frac{1}{2}}(\lambda)=2cos(\pi NP).
$$
In the case when $p=\frac{l+1}{N}$, where $l\ge 0$, $l\in \mathbb{Z}$ there 
exists an additional number of truncations:
$$
\mathbf{t}_{\frac{N}{2}-\frac{1}{4}}(\lambda)=0, 
\quad \mathbf{t}_{\frac{N}{2}}(\lambda)=
\mathbf{t}_{\frac{N}{2}-\frac{1}{2}}(\lambda)=(-1)^{l+1},
$$
$$
\mathbf{t}_{\frac{N}{2}-\frac{1}{2}-s}(\lambda q^{\frac{N}{2}})=
(-1)^{4s}\mathbf{t}_{s}(\lambda)(-1)^{l+1}.
$$
These relations allow us to rewrite the fusion relation system
in the Thermodynamic Bethe Ansatz Equations of $D_{2N}$ type.
\section{conclusions}
In this paper we studied algebraic relations arising from the integrable 
structure of CFT provided by the SUSY $N$=1 KdV hierarchy. 
The construction of the $\mathbf{Q}$-operator as 
a ``transfer''-matrix corresponding to the infinite-dimensional 
q-oscillator representation could be also applied to the lattice models.
 The relations like Baxter's and fusion ones 
will be also valid
in the lattice case because they depend only on the decomposition properties 
of the representations.\\ 
\hspace*{5mm} In the following we also plan to study the quantization of 
$N>1$ SUSY KdV hierarchies, related with super-W conformal/topological 
integrable field theories.

\begin{acknowledgments}
I am very grateful to my supervisor Prof. P. Kulish. It is a pleasure to thank 
Prof. M. Semenov-Tian-Shansky and F. Smirnov 
for useful discussions and Prof. L. Baulieu, B. Pioline and 
LPTHE, Univ. Paris 6 for support and hospitality.
This work was supported by Dynasty Foundation and CRDF grant RUM1-2622-ST-04.
\end{acknowledgments}

\end{document}